\documentclass{aa}  
\usepackage{graphicx}
\usepackage{txfonts}
\usepackage{nicefrac}
\usepackage{subfig} 
\newcommand{\teff}{\mbox{$T_{\rm eff}$}}
\newcommand{\logg}{\mbox{$\log{g}$}}
\newcommand{\mh}{\mbox{[M/H]}}
\newcommand{\feh}{\mbox{[Fe/H]}}
\newcommand{\afe}{\mbox{[$\alpha$/Fe]}}
\newcommand{\vrad}{\mbox{$v_{\rm rad}$}}
\newcommand{\Msun}{\mbox{$M_{\odot}$}}

\newcommand{\phx}{PHOENIX}
\newcommand{\muse}{MUSE}

\defcitealias{2015A&A...subm....K}{Paper~II}

\begin{document} 

   \title{\muse\ crowded field 3D spectroscopy of over 12\,000 stars in the globular cluster \object{NGC~6397}}
   \subtitle{I. The first comprehensive HRD of a globular cluster\thanks{Data products are available at http://muse-vlt.eu/science.}$^,$\thanks{Based on observations obtained at the Very Large Telescope (VLT) of the European Southern Observatory, Paranal, Chile (ESO Programme ID 60.A-9100(C)).}}

   \author{Tim-Oliver~Husser\inst{1}, 
           Sebastian~Kamann\inst{1}, 
           Stefan~Dreizler\inst{1}, 
           Martin~Wendt\inst{2,3},
           Nina~Wulff\inst{1},
           Roland~Bacon\inst{4},
           Lutz~Wisotzki\inst{2},
           Jarle~Brinchmann\inst{5},
           Peter~M.~Weilbacher\inst{2}, 
           Martin~M.~Roth\inst{2},
           Ana~Monreal-Ibero\inst{6}}
   
   \authorrunning{Husser et al.}

   \institute{$^1$ Institut f\"ur Astrophysik, Georg-August-Universit\"at G\"ottingen, Friedrich-Hund-Platz 1, 37077 G\"ottingen, Germany\\
	      $^2$ Leibniz-Institut für Astrophysik Potsdam (AIP), An der Sternwarte 16, D-14482 Potsdam, Germany\\
	      $^3$ Institut f\"ur Physik und Astronomie, Universit\"at Potsdam, 14476 Potsdam, Germany\\
	      $^4$ CRAL, Observatoire de Lyon, CNRS, Université Lyon 1, 9 avenue Ch. André, 69561 Saint Genis-Laval Cedex, France\\
	      $^5$ Leiden Observatory, Leiden University, PO Box 9513, 2300 RA Leiden, The Netherlands\\
	      $^6$ GEPI, Observatoire de Paris, CNRS, Université Paris-Diderot, Place Jules Janssen, 92190 Meudon, France}

   \date{Received -; accepted -}
 
  \abstract
   {}
   {We demonstrate the high multiplex advantage of crowded field 3D spectroscopy using the new integral field spectrograph MUSE by means of a spectroscopic analysis of more than 12\,000 individual stars in the globular cluster NGC 6397.}
   {The stars are deblended with a PSF fitting technique, using a photometric reference catalogue from HST as prior, including relative positions and brightnesses. This catalogue is also used for a first analysis of the extracted spectra, followed by an automatic in-depth analysis using a full-spectrum fitting method based on a large grid of \phx\ spectra.}
   {With $18\,932$ spectra from $12\,307$ stars in \object{NGC~6397} we have analysed the largest sample so far available for a single globular cluster. We derived a mean radial velocity of \vrad=17.84$\pm$0.07\,km\,s$^{-1}$ and a mean metallicity of \feh=-2.120$\pm$0.002, with the latter seemingly varying with temperature for stars on the RGB. We determine \teff\ and \feh\ from the spectra, and \logg\ from HST photometry. This is the first very comprehensive HRD for a globular cluster based on the analysis of several thousands of stellar spectra, ranging from the main sequence to the tip of the RGB. Furthermore, two interesting objects were identified with one being a post-AGB star and the other a possible millisecond-pulsar companion.}
   {}


   \maketitle
%
%________________________________________________________________

\section{Introduction}
\begin{figure*}
 \includegraphics[width=18cm]{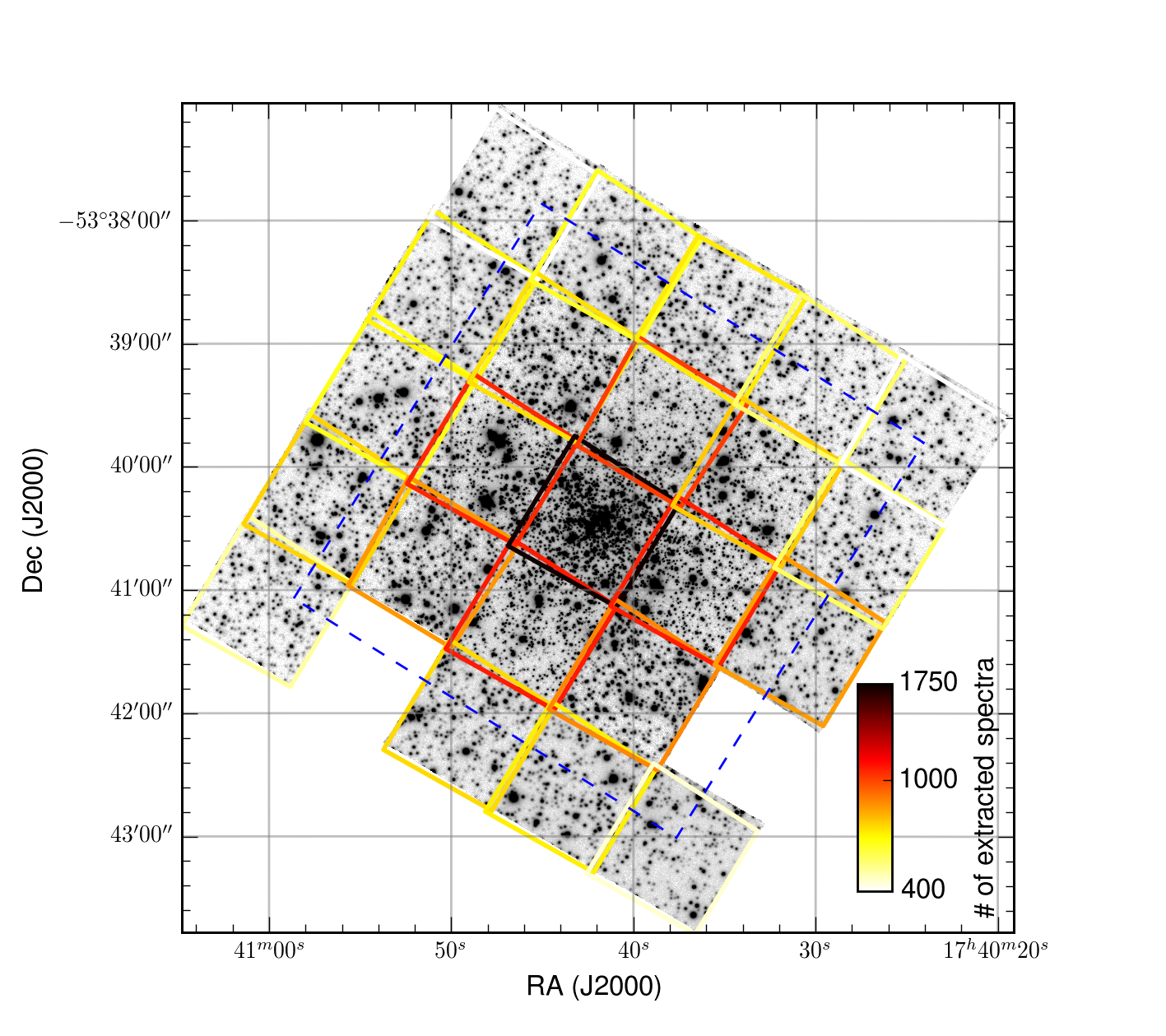}
 \caption{This V band image was extracted from the mosaic of \muse\ data cubes of \object{NGC~6397}. Overplotted are the single pointings, which have been observed multiple times at different position angles and with little offsets. Two of the planned pointings were not carried out, so the mosaic is incomplete. The colour coding indicates the number of stellar spectra that were extracted from each pointing, ranging from $\sim$1\,700 in the central pointing to about 400 in the outer fields. The dashed blue line indicates the area covered by the ACS survey of Galactic globular clusters \citep{2007AJ....133.1658S}.}
 \label{fig:obs:mosaic}
\end{figure*}
Over the last decades, globular cluster research has mainly been dominated by two complementary approaches. On the one hand, photometry at high spatial resolution, mainly using the \emph{Hubble} space telescope (HST), aims to study the populations of the clusters as a whole, down to the faintest stars \citep[e.g.][]{2007AJ....133.1658S}. On the other hand, spectroscopy at high spectral resolution tries to infer cluster properties from detailed studies of individual stars, mainly giants \citep[e.g.][]{2009A&A...505..117C}. Both approaches were able to show that the stellar populations in the clusters are not as homogeneous as once thought. The photometrically detected splits in the main sequences of some massive clusters \citep[see e.g.,][]{2007ApJ...661L..53P,2010AJ....140..631B} or the more common variations in the abundances of light elements found in high-resolution spectroscopic studies \citep[for a review see][]{2012A&ARv..20...50G} are famous examples in this respect.

These findings raised new questions. The variations in the light element abundances are often explained by a second generation of star formation, from material polluted by the first stellar generation. However, the nature of the polluters is still debated \citep[see, e.g.,][]{2001ApJ...550L..65V,2006A&A...458..135P,2009A&A...507L...1D}. Alternatively, early disc accretion has been proposed by \citet{2013MNRAS.436.2398B} to explain the variations without a second star formation period.

An observational approach that has not been followed so extensively compared to the two aforementioned ones is spectroscopy at low to medium resolution (with R$\sim$3\,000). Obviously, it is unlikely to achieve the same accuracy in element abundances as high resolution spectroscopy, but it has other advantages: the lower resolution usually comes with a longer wavelength range and shorter exposure times. The potential of applying full-spectrum fitting to medium resolution spectra has been demonstrated, among others, by \cite{2009A&A...501.1269K} and \cite{2008ApJ...682.1217K}.

A common problem of spectroscopic studies is that only a limited number of stars in a globular cluster can be meaningfully observed. The catalogues of the ACS survey of Galactic globular clusters \citep{2008AJ....135.2055A} lists about 23\,000 stars in just the central 1\,arcmin$^2$ of 47 Tuc (NGC 104), so blending is a common problem, usually limiting spectroscopy to brighter stars or outer regions of the clusters.

After first attempts with longslit spectroscopy, resulting in only handfuls of spectra of bright isolated giants \citep[see, e.g.,][]{1959MNRAS.119..157K}, multi-object spectroscopy (MOS) took over, increasing the number of observed objects steadily. For instance, \cite{2009A&A...493..947S} managed to obtain 5\,973 individual high-resolution spectra of 2\,469 stars in M4 using the FLAMES+GIRAFE facility at the VLT. Still, this technique suffers from contamination from nearby -- especially brighter -- stars. Hence, MOS studies typically target giant stars in the outskirts of the clusters.

An important advantage of photometric studies has consisted in the ability to use the full spatial information within the field of view for deblending techniques, e.g. PSF fitting tools like \texttt{daophot} \citep{1987PASP...99..191S}. An approach usable for spectroscopy, for instance employed by \cite{1995AJ....110.1699G}, was scanning Fabry-Perot (FP) spectroscopy, giving full spatial information and thus allowing for efficient PSF deblending. The drawbacks are a small wavelength range and especially the risk of changing environmental properties like seeing and sky brightness, which complicates the analysis.
\begin{figure*}
 \includegraphics{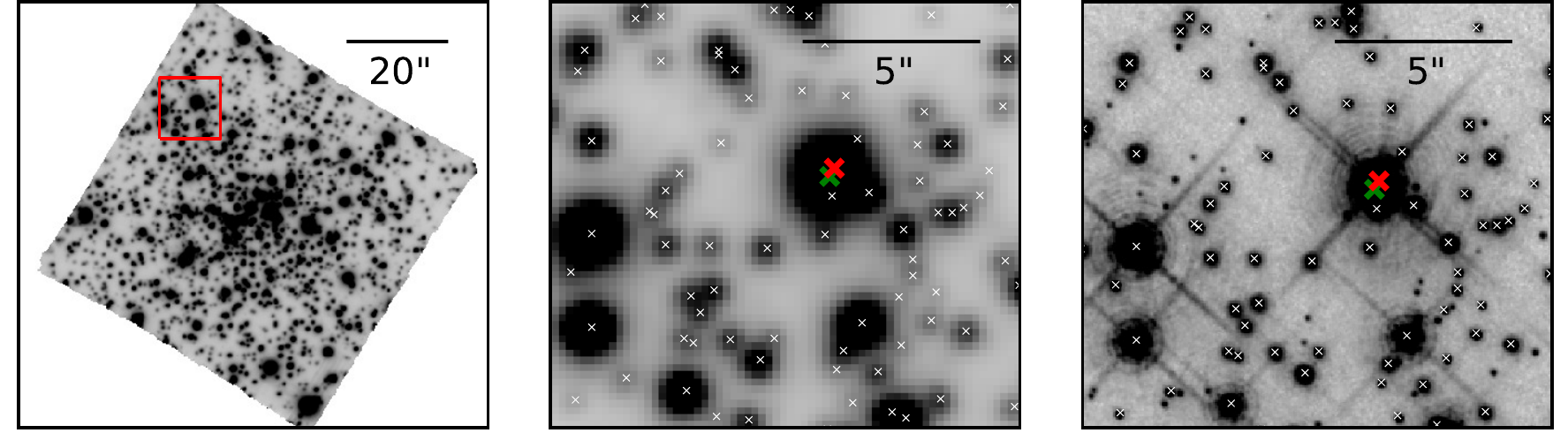}\\
 \includegraphics{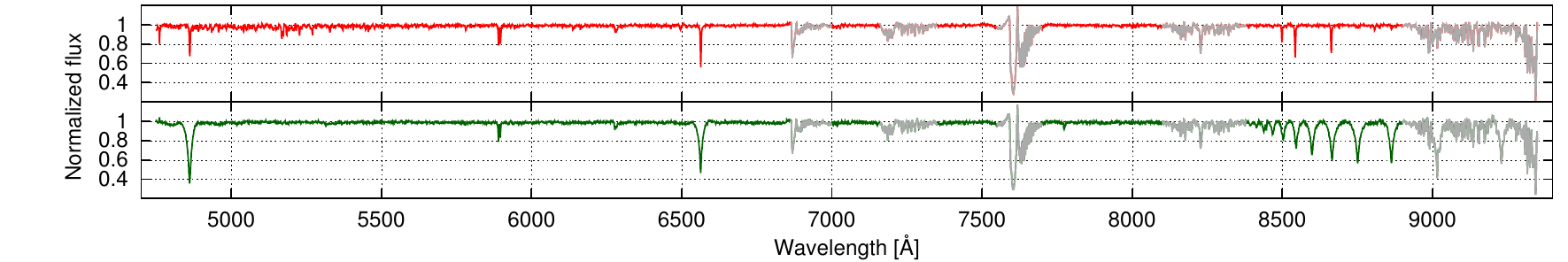}
 \caption{Deblending of stellar spectra on \muse\ data. The top left panel shows a whitelight image of the central \muse\ datacube in \object{NGC~6397}. A zoom into the area marked by a red square is shown in the top middle panel. In the top right panel, the same field of view is shown as seen with HST. All stars for which spectra were extracted are marked by crosses. Spectra extracted from the stars marked with the red and green crosses are shown in the lower panel and show clearly distinct features without any contamination from each other. The star marked in red was determined to be a late type G star, while the green one is of type A.} Note that they have been continuum subtracted and normalised and that areas dominated by telluric absorption have been marked in grey for clarity.
 \label{fig:extract:example}
\end{figure*}

Integral field spectrographs (IFS) that also preserve the full spatial information provide a solution to this shortcoming, yet the limiting factor of these instruments has always been the small number of spatial pixels which made the observation of large stellar samples very time consuming. However, the recently commissioned IFS \muse\ \citep[Bacon et al. in prep., see also][]{2010SPIE.7735E..08B,2014Msngr.157...13B}, mounted at UT4 (Yepun) of the ESO Very Large Telescope (VLT) changes this. It provides an unprecedented combination of a large field of view ($1\arcmin\times1\arcmin$) with a fine spatial sampling ($0\farcs2$). Dedicated simulations by \citet{2013A&A...549A..71K} predict that up to $5\,000$ stars can be simultaneously observed in a single \muse\ pointing. This number would already exceed the largest spectroscopic samples obtained so far. In contrast to previous studies, \muse\ observations are not limited to isolated stars but promise to encompass the complete stellar population of a cluster in a single observation and therefore to bridge the gap to photometric studies.

In this paper, we discuss \muse\ observations of the globular cluster \object{NGC~6397} which is an extremely metal-poor, core-collapsed globular cluster with a moderate stellar mass. Thanks to being one of the closest clusters with a distance of only 2.4\,kpc \citep{1996AJ....112.1487H}, it has been subject of spectroscopic and photometric studies. This makes it a perfect object for testing the capabilities of \muse\ in the analysis of globular cluster populations.

The paper is structured as follows. In Sect.~\ref{sec:obs} the observations and the basic data reduction will be discussed, following with the source extraction in Sect.~\ref{sec:extract}. In Sect.~\ref{sec:analysis} we will describe our full-spectrum fitting technique used for analysing the obtained spectra. The results are presented in Sect.~\ref{sec:results}, including a comprehensive Hertzsprung-Russell diagram (HRD), a peculiar trend in metallicity and some interesting individual objects. The conclusions follow in Sect.~\ref{sec:conclusions}. A more in-depth analysis of the kinematics and possible evidence for a black hole in the centre of \object{NGC 6397} are discussed in \citet[][hereafter \citetalias{2015A&A...subm....K}]{2015A&A...subm....K}.

\section{Observations and data reduction}
\label{sec:obs}
In order to explore the capabilities of \muse\ in crowded stellar fields, the globular cluster \object{NGC~6397} was observed during \muse\ commissioning, lasting from July 26nd to August 3rd, 2014. We created a $5\times5$ mosaic of the central part of the cluster, reaching out to a distance of $\sim3\farcm5$ from the cluster centre. The various fields of the mosaic are shown in Fig.~\ref{fig:obs:mosaic}. The reason for the two missing fields in the outer rim is that the observations were embedded into the commissioning plan and therefore interrupted occasionally to perform instrument tests.
 
Each pointing was observed with a dither pattern and offsetting the derotator by 90 degrees between exposures. This was done to ensure that the light of each star gets dispersed by multiple spectrographs which improves the flatfielding and hence yields a homogeneous image quality across the field of view. The individual exposure times were always short, $\leq60\,\mathrm{s}$, to avoid saturation of the brightest cluster giants. In total, we obtained $127$ exposures with a total integration time of $95\,\mathrm{min}$. Including overheads, the observations took about $6\,\mathrm{h}$.

The seeing during the observations was a bit variable, but never worse than $\sim1\arcsec$. However, the observations of the central part of the cluster were carried out under excellent seeing conditions of $\sim0\farcs6$.
 
The data reduction was done using the official \muse\ pipeline (in versions 0.18.1, 0.92, and 1.0) written by Weilbacher et al. \citep[in prep., see also][]{2012SPIE.8451E..0BW}. It performs the basic reduction steps -- bias subtraction, flat fielding, and wavelength calibration -- on an IFU-per-IFU basis. In the next step, the data of each IFU was transformed into spatially calibrated coordinates before the data from all 24 IFUs were combined into a single pixtable, a FITS-file containing for each CCD pixel information about its flux and uncertainty, wavelength, and spatial position inside the field of view. To combine the pixtables from the individual exposures of each pointing into a single datacube, we had to account for the small offsets that occur during position angle changes, because the rotation centre is not perfectly aligned with the optical axis of the instrument (``derotator wobble''). We did so by creating a broadband image from each pixtable, measuring the coordinates of the brighter stars in the field of view, and feeding the measured average offsets to the pipeline when creating the final datacube for each pointing.
 
For visualisation purposes, we also created a datacube from the entire mosaic, but in a narrow wavelength range around 6\,000\,\AA. A collapsed image is shown in Fig.~\ref{fig:obs:mosaic} and illustrates the remarkable image quality that is achieved by \muse.

\section{Extraction of stellar spectra}
\label{sec:extract}
To obtain clean spectra in a crowded region such as a globular cluster, sophisticated analysis techniques are required. We developed a code to perform source deblending via PSF fitting on IFS data, described in detail in \cite{2013A&A...549A..71K}. It relies on the existence of a photometric reference catalogue, including relative positions and brightnesses for the stars in the field of view. For NGC~6397, we used the catalogue prepared by \citet{2008AJ....135.2055A} as part of the ACS survey of Galactic globular clusters \citep{2007AJ....133.1658S}. Our \muse\ mosaic extends beyond the ACS observations in this cluster (see Fig.\ref{fig:obs:mosaic}), therefore we had to obtain positions and brightnesses of stars there by other means. To this aim, we created broadband images from the datacubes of these fields and analysed them with \texttt{daophot} \citep{1987PASP...99..191S}. Afterwards, we identified the stars  that exist in the ACS catalogue, used those to calibrate the magnitudes measured with \texttt{daophot}, and added the missing stars to the catalogue. A possible shortcoming of this approach is that our deblending code works best with a reference catalogue obtained at higher spatial resolution while the broadband images obviously have the same resolution as the IFS data. However, the areas in \object{NGC~6397} without coverage by the ACS catalogue are not that densely crowded. Therefore we do not expect a significant impact on the results of the spectrum extraction.
 
For details on the method we refer the reader to \cite{2013A&A...549A..71K}, here follows a brief summary of the methodology. The extraction of the stellar spectra is performed in several steps. The analysis is started from an initial guess of the PSF in the \muse\ data, which we model as an analytic Moffat profile with up to four free parameters -- the $\mathrm{FWHM}$, the kurtosis $\beta$, the ellipticity $e$, and the position angle $\theta$, all of which can be wavelength dependent. For the initial guess, we tailor the $\mathrm{FWHM}$ to the seeing of the observation and set $\beta$ to an average value of $2.5$. The \muse\ PSF did not show any signs of ellipticity, hence we set $e=0$. In the next step, the PSF model is used to create a mock \muse\ image from the reference catalogue. By cross-correlating this image against the \muse\ cube, an initial guess for the coordinate transformation is obtained. For this purpose, we employ an affine transformation with up to six free parameters. Then, the sources for which meaningful spectra can be extracted are identified. To this aim, the signal-to-noise ratio (S/N) of each source is predicted based on its magnitude in the reference catalogue, the PSF, and the variances of the \muse\ data. In addition, the density of brighter sources around the source in consideration is determined. Only sources that pass a S/N cut of $5$ and for which the density calculation yields $<0.4$ brighter sources per resolution element are used in the further analysis. This set of sources is then used in the actual extraction process, which works iteratively on every layer, i.e.\ every wavelength slice, of the datacube. In each layer the analysis starts with a simultaneous fit to the fluxes of all sources, using the current estimate of the PSF and the source coordinates. Afterwards, all sources except those identified as isolated enough to model the PSF are subtracted and the parameters of the PSF and the coordinate transformation are refined. The new estimates are then used in another simultaneous flux fit. This process is iterated until convergence is reached on the source fluxes and the analysis of an adjacent layer is started. In this layer, the final estimates of the PSF and the coordinates are used as initial guesses.

\begin{figure}[t!]
 \includegraphics{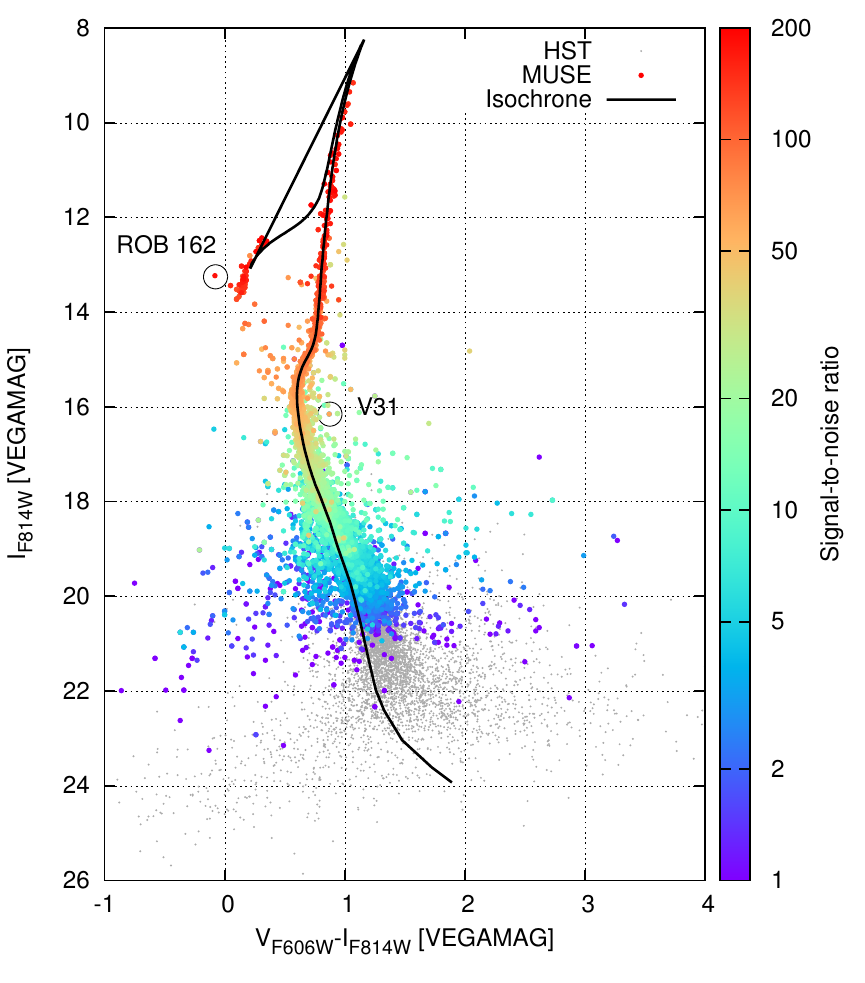}
 \caption{CMD of \object{NGC~6397} generated from the input catalogue with colour-coded S/N ratios of extracted spectra. The two stars marked with circles will be discussed in Sections~\ref{sec:sdo} and \ref{sec:pulsar}. Overplotted in black is an isochrone with an age of 13.5\,Gyrs and Z=0.0014 (i.e.\ \feh$\approx$-2.13).}
 \label{figure:cmd}
\end{figure}
After all layers have been processed in this way, a final PSF model is derived for the whole datacube. To this aim, the values of the PSF parameters obtained in the individual layers are fitted with polynomials. Polynomial fits are also performed to the source coordinates measured across the cube. The reason why this is done is to reduce the effect of small random jumps between layers and thereby to increase the S/N. Using polynomials for this task is justified because ambient characteristics like atmospheric refraction or the seeing should result in a smooth change of the PSF and the source coordinates with wavelength. Indeed we found that the $\mathrm{FWHM}$ always shows a monotonic decrease with wavelength, very close to the theoretically expected behaviour. On the other hand, $\beta$ did not vary strongly with wavelength.

In the last step, the final spectra are extracted by going through all layers of the cube once more and using the new estimates of the PSF and the source coordinates to extract the fluxes. This was done again by a simultaneous flux fit to all sources.

The results of the source extraction are summarised in Figs.~\ref{fig:extract:example} and \ref{figure:cmd}. Figure~\ref{fig:extract:example} gives a good idea of the power of deblending techniques in integral field data. It allows to obtain clean spectra even in cases of heavily blended stars, such as the two stars marked by coloured crosses in Fig.~\ref{fig:extract:example}. Based on the magnitudes in the reference catalogue, one is a red giant (indicated by a red cross in Fig.~\ref{fig:extract:example}) while the other is a horizontal branch star. The spectra we extracted and show in Fig.~\ref{fig:extract:example} nicely confirm this classification. Broad Paschen lines are visible in the red part of the spectrum of the horizontal branch star while the red giant shows the characteristic calcium triplet lines in the same spectral window. Neither spectrum shows any contamination caused by the close neighbour. Also we can reproduce the $V-I$ colours of the reference catalogue to $\sim0.1\,\mathrm{mag}$ accuracy.

In Fig.~\ref{figure:cmd} we show a colour-magnitude diagram (CMD) generated from the input catalogue where we colour coded the stars by the S/N of the extracted spectra. Our spectra cover a wide range of stellar types, ranging from red giants over horizontal branch stars and blue stragglers to main sequence (MS) stars well below the turn-off (TO). Recall that the total exposure times per pointing  were $4\,\mathrm{min}$ at most, but thanks to the high throughput of \muse\ and our extraction method which maximises the S/N of every extracted spectrum, we are still able to observe stars as faint as $\mathrm{I}\approx20$ at a S/N of $\sim$5.
 
\begin{figure}[t!]
 \includegraphics{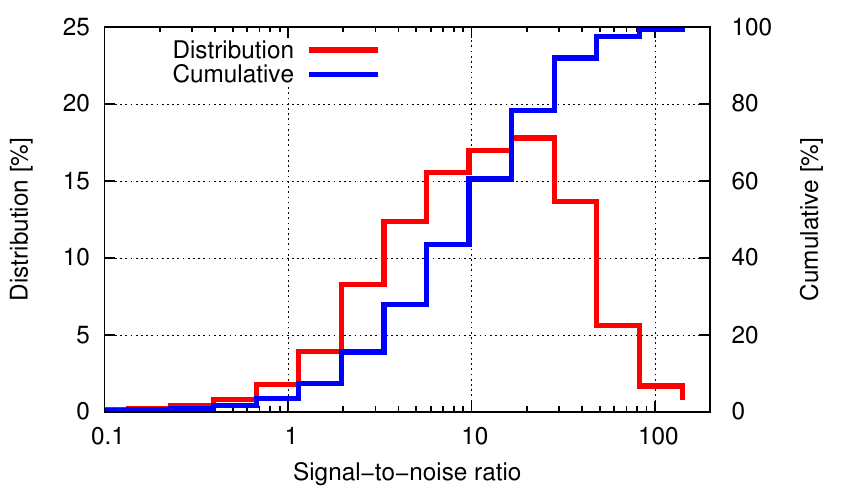}
 \caption{The distribution of S/N ratios for the extracted spectra is plotted in red. Overplotted in blue is the cumulative distribution, showing that about 50\% of the spectra have S/N>10.}
 \label{figure:snr}
\end{figure}

In total, we were able to extract $18\,932$ spectra, $10\,521$ ($\sim$55.6\%) of which have a S/N>10, for $12\,307$ stars, i.e.\ some have been observed multiple times in overlapping pointings. The distribution of S/N ratios is plotted in Fig.~\ref{figure:snr}. Taking only those stars into account that we determined to be cluster members (see Sect.~\ref{section:vrad}), we obtained spectra of 894 stars (13\%) that are brighter than the turn-off (assumed to be at 15.7\,mag in F814W) and of 6\,248 stars (87\%) that are fainter, i.e.\ on the main-sequence. To our knowledge, this is already the largest spectroscopic sample ever obtained for an individual cluster, an impressive demonstration of the unprecedented multiplexing capabilities of \muse.

\subsection{Line spread function}
\label{section:lsf}
\begin{figure}[t!]
  \includegraphics{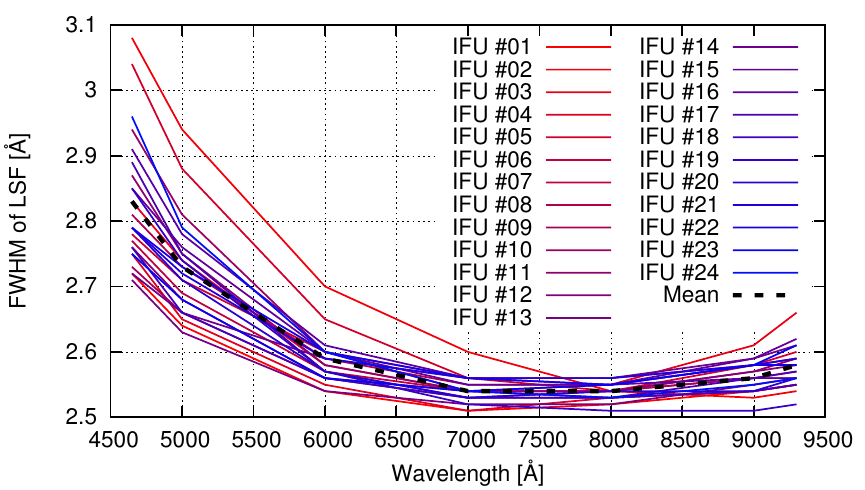}
  \caption{Full widths at half maximum of the line spread functions of all 24 IFUs of \muse\ at different wavelengths as taken from a single exposure.}
  \label{figure:lsf}
\end{figure}

In a low-to-medium resolution spectrograph like \muse, the knowledge of the line spread function (LSF) is quite important, since its shape has a stronger influence on the observed spectra than with high-resolution spectrographs. Each of the 24 IFUs of \muse\ has slightly different characteristics and the LSF of each is highly wavelength dependent. The \muse\ data reduction system tries to estimate those profiles -- as shown in Fig.~\ref{figure:lsf} for a single exposure -- for each IFU at different wavelengths by fitting sky lines with Gauss-Hermite polynomials.

As one can see, the profiles change significantly both with wavelength and with the IFU. While the former variation can be dealt with quite easily using step-wise convolution, the latter poses a more serious problem, since the flux measured for a single star can be distributed over several IFUs. In that case, picking an appropriate LSF is not an obvious choice. For the present analysis we decided to use a mean LSF (see dashed line in Fig.~\ref{figure:lsf}), which will overestimate the spectral resolution for some IFUs and underestimate it for others. Except for two IFUs, the deviation from this mean is never larger than 0.13\,\AA\ and usually well below 0.05\,\AA. 

The approach of using a mean LSF is further justified by the fact that we ran a dither pattern on each observed pointing by rotating 90 degrees between exposures. This way the spectrum of a single star is a combined measurement of multiple spectrographs which, in turn, should have an LSF close to the mean one that we are using.

We pre-convolve all our model spectra with a kernel $K$ that is slightly higher resolved than the LSF (with an offset $\sigma_\mathrm{off}$), i.e. $\sigma_K=\sqrt{\sigma_\mathrm{LSF}^2-\sigma_\mathrm{off}^2}$. This allows for some additional Gaussian line broadening during the fitting process (see Section \ref{subsection:fitting}), which should approximately compensate for the mean LSF that we are using and allow for some intrinsic line broadening. 

For testing this, we ran the analysis (as described in the next section) for the central pointing with two more grids of model spectra that have been convolved with different LSFs, namely those of IFUs \#2 and \#13 which correspond to the highest and the lowest curve in Fig.~\ref{figure:lsf} respectively. We could not find any systematic offsets and the scatter between the results was small, i.e.\ $\sigma_{\mh}=0.06$\,dex for the metallicity and $\sigma_{\teff}=6$\,K for the effective temperature for S/N>20, which we consider to be the lower limit for reliable results (see Table~\ref{table:uncertainties}). And as expected, the additional line broadening applied to the model spectra is smallest for those convolved with the LSF of IFU \#2, i.e.\ the broadest one.

Although we already obtain reasonable results with this model for the LSF, we have to state that it still has space for improvements. For instance, the LSF of \muse\ appears to be slightly asymmetric, which we do not take into account at all at the moment. Furthermore we are currently ignoring the Hermite part of the LSF as given by the pipeline. Due to this, currently we do not fit the surface gravity -- for which the fit highly depends on the line shapes and therefore on the LSF -- spectroscopically, but use a value derived from the comparison with an isochrone.

\section{Analysis}
\label{sec:analysis}
Our analysis of the observed spectra is a multi-step process that will be explained in detail in this section. The steps involved are:
\begin{enumerate}
 \item We compare the HST photometry with an isochrone with age, metallicity, and distance taken from the literature (with some minor manual adjustments for a better match), resulting in initial guesses for \teff\ and \logg.
 \item We obtain a model spectrum with given \teff/\logg/\mh\ and do a cross-correlations for each observed spectrum for deriving radial velocities.
 \item We run an optimisation with \teff, \mh, \vrad, line broadening and telluric absorption as free parameters, fixing \logg\ to the photometric value, i.e.\ assuming the stars belongs to the observed stellar cluster.
 \item We optimise again with \feh\ and \afe\ as free parameters, fixing the others to the previously obtained values.
\end{enumerate}
This procedure yields a list of parameters for all the stars in the sample. For most of the further analyses of the results a S/N cut of 20 is applied.

\subsection{Full-spectrum fitting}
\label{subsection:fitting}
The method we use for determining stellar parameters is similar to the one described by \cite{2009A&A...501.1269K} and was already discussed briefly in \cite{2012ASInC...6...71H}. Instead of fitting only parts of an observed spectrum, e.g.\ strong absorption features, we always use the whole available wavelength range, and mask, if at all, only small regions with known template mismatches. During the fit, the observed spectrum $O(x)$, given in pixel space, is never transformed in any way, i.e.\ convolved or rebinned. Instead, we try to find a best-fit model $G$ as the weighted sum of model spectra $M_k$ with coefficients $\omega_k$, each convolved with a corresponding kernel $B_k$:
\begin{equation}
 G = \sum\limits_{k=1}^{K} \omega_k \left[ B_k \ast M_k \right].
\end{equation}
We perform the convolution (denoted by $\ast$) in the logarithmic domain, which allows us to apply both line shifts (i.e.\ radial velocities) and broadenings in one single step. As the sum suggests, we can fit multiple models to a single observed spectrum, but for this work we always used $K=1$, i.e.\ a single stellar component is fitted -- which reduces the problem to finding a single model $M_1$ with the convolution kernel $B_1$ that matches the observation best. In case a spectrum is known to be a combination of multiple stars, $K$ can be increased for determining the parameters of all stars.

The best-matching model is then found by running a Levenberg-Marquardt optimisation. At each iteration a model spectrum is interpolated from a given spectral library (see next section). This is then convolved by a kernel that both broadens the lines by a given amount and shifts the spectrum to a given radial velocity. Finally the model spectrum is rebinned to the same wavelength grid as the observed spectrum. 

The model spectra that we use are usually not normalised and as mentioned before we never alter the observed spectrum in any way, so they still contain their continuum fluxes. Instead we fit a Legendre polynomial, which, multiplied with the model, best matches the observation. This is done in every iteration and for every interpolated model spectrum, so we always find the best polynomial to match the two spectra. The order of the polynomial is chosen in a way that stellar features are not mistreated as continuum.

The free parameters in the fit are typically the effective temperature $\teff$ and the metallicity \mh. Furthermore we allow for a radial velocity $v_\mathrm{rad}$ and an intrinsic line broadening using a Gaussian with the standard deviation $\sigma$. In general we can also fit the surface gravity \logg, but due to our current problems with understanding the LSF (see Sect.~\ref{section:lsf}) this has not been done for the analyses presented in this publication.

As stated before, we usually measure the metallicity as a scaled solar abundance \mh. Since most globular clusters have been found to be over-abundant in alpha elements, this value changes when adding \afe\ as an additional free parameter to the fit. In that case we denote the metallicity as \feh\ in order to be able to distinguish between both.

The convolution with a Gaussian kernel that we apply to the model spectra also allows us to account for effects like stellar rotation and macro-turbulence. The former is actually a non-Gaussian broadening, but at the spectral resolution of \muse\ the difference is small. Furthermore, in globular clusters we expect only very old stars with relatively small masses ($<1\,\Msun$), so we do not expect to observe large rotational velocities.

We tested the fitting procedure extensively against other publicly available medium-resolution stellar libraries, both synthetic and empirical. Among those were \textit{The Indo-U.S. Library of Coudé Feed Stellar Spectra} \citep[also known as \textit{cflib},][]{2004ApJS..152..251V}, \textit{ELODIE} \citep{2001A&A...369.1048P}, and \textit{SpectraLib} \citep{2005A&A...442.1127M}. The results were all in good agreement and no systematic errors were found.

\subsection{Model spectra and interpolation}
\renewcommand{\tabcolsep}{0.5mm}
\begin{table}
  \caption{Parameter space of the \phx\ grid of model spectra used for fitting the observations. For $-3.0<\feh<0.0$ the available alpha element abundance are limited to $\afe \in [-0.4,0.0,+0.4,+1.2]$ and for those remaining the effective temperature \teff\ is restricted to 3\,500 -- 7\,000K.}
  \label{table:paramspace}
  \centering                                      %
  \begin{tabular}{lrrlc}
    \hline \hline
    Variable	& & \multicolumn{2}{c}{Range}  & Step size \\ \hline
    $\teff$ [K] & & 2\,300 & -- 7\,000         & 100 \\
		& & 7\,000 & -- 12\,000        & 200 \\
		& &12\,000 & -- 15\,000        & 500 \\
    $\logg$     & &    0.0 & -- +6.0           & 0.5 \\
    $\feh$      & &   -4.0 & -- -2.0           & 1.0 \\
		& &   -2.0 & -- +1.0           & 0.5 \\
    $\afe$      & &   -0.2 & -- +1.2           & 0.2 \\ \hline
  \end{tabular}
\end{table}
\renewcommand{\tabcolsep}{2mm}
For analysing the observed spectra we use an extended version of the \emph{G\"ottingen Spectral Library} \citep{2013A&A...553A...6H}. We increased the upper limit for the effective temperature \teff\ to 15\,000\,K and added some variations in alpha element abundances for the metallicities that were previously missing.

When creating the spectral library we used values for the micro-turbulence $v_\mathrm{micro}$ empirically derived from 3D radiative hydrodynamic models models \citep[see][]{2009A&A...508.1429W}. This was done for M stars only, but the extrapolation towards higher temperatures seemed plausible. This is a different approach than the commonly seen one of using a fixed value for the micro-turbulence. \cite{2015ApJ...804..113B} suggested that there might be a $v_\mathrm{micro}$-\feh\ degeneracy, which could affect our results -- but the same would be true with a fixed micro-turbulence.

Since with the Levenberg-Marquardt algorithm we use an optimisation routine that depends on gradients, interpolating linearly within the model grid is not an option, because that would produce discontinuities in the first derivatives. Instead, we use a scheme based on cubic splines. The interpolation is done under the assumption that each wavelength point is independent, so that we can interpolate them individually.

In a single dimension $p$ of the grid, we derive second derivatives (along this dimension) by fitting a spline to the fluxes. With $p_0$ being the value of the parameter to interpolate and $p_1$ and $p_2$ being the two adjacent points in the grid, we can calculate the following coefficients \citep[see][]{Press:2007:NRE:1403886}:
\begin{align}
  A &= \frac{p_2 - p_0}{p_2 - p_1}, &B &= 1 - A,\\
  C &= \nicefrac{1}{6} A^4 (p_2 - p_1)^2, &D &= \nicefrac{1}{6} B^4 (p_2 - p_1)^2,
\end{align}
which in turn we can use to interpolate the flux $f$ at $p_0$ from the fluxes and second derivatives (along $p$) at $p_1$ and $p_2$:
\begin{equation}
  f_\lambda(p_0) = A f_\lambda(p_1) + B f_\lambda(p_2) + C f_\lambda''(p_1) + D f_\lambda''(p_2) .
\end{equation}

Applying this interpolation scheme to higher dimensions can easily be done by running it recursively, i.e.\ interpolating in one dimension after the other. By precomputing the second derivatives in one dimension (in our case \teff) we can reduce the computation time for the interpolation significantly.

\subsection{Initial guesses}
The performance of optimisation algorithms like Levenberg-Marquardt usually depends on good initial guesses. Therefore we did comprehensive tests with varying starting parameters. We found that the results for the radial velocity are quite insensitive to its initial guess, provided that absorption lines in observation and model still overlap -- which usually is the case for $\pm200$\,km/s around the real value. A similar result was observed for the effective temperature. Varying its initial guess across the full range of the used spectral library did not affect the results at all. For the metallicity we found that the real value could not be recovered reliably when starting with a low value like, e.g.\ \mh=-4.0. But this was to be expected due to the low number of absorption lines in those spectra. Thus, the initial guess for \mh\ is set to the mean cluster metallicity throughout this work.

Since we already used a catalogue including brightnesses in different passbands for the extraction of the spectra, we can use it again for a comparison with a matching isochrone with an age of 13.5\,Gyrs and Z=0.0014 ($\feh\approx-2.13$) from \cite{2012MNRAS.427..127B}\footnote{\url{http://stev.oapd.inaf.it/cmd}}, as shown in Fig.~\ref{figure:cmd}. We fit a two-dimensional polynomial $P(V, V-I)$ of third order to the effective temperatures from the isochrone as function of colour and magnitude. Evaluating this polynomial for each star in the catalogue yields initial guesses for the effective temperature. The same procedure is applied for the surface gravity. We estimate initial guesses for each star in the CMD by simply evaluating those two polynomials. The values obtained for \teff\ and \logg\ are then used to create a template spectrum for a cross-correlation, which yields the initial radial velocity for the full-spectrum fit.

Please note that this method will most likely produce wrong \logg\ values for field stars, since they are obtained by comparison with an isochrone with most probably wrong age, metallicity, and distance. As discussed above, we are currently fixing the surface gravity to this initial guess, for these stars we will see an impact on the results for \teff\ and \mh. As will be shown in Paper~II, the measured metallicities of the field stars in the sample are distributed as expected by a Milky Way model and therefore we expect the effect on \mh\ to be small. As our membership determination depends on metallicities and radial velocities -- which are not affected by this --, we can effectively remove such stars from further analysis.

\subsection{Telluric absorption}
\begin{figure}[t!]
  \includegraphics[width=8.8cm]{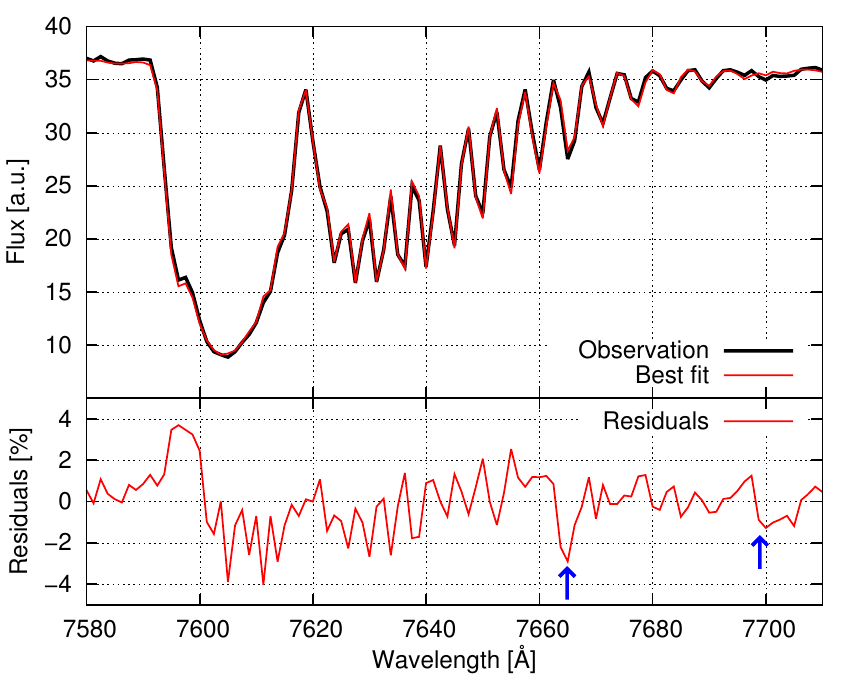}
  \caption{Fraunhofer A band and the best-fitting model in the spectrum of a typical G subgiant in our sample. Marked with blue arrows in the residuals are the positions of two \ion{K}{i} absorption lines from the interstellar medium.}
  \label{figure:tellfit}
\end{figure}
The concept of full-spectrum fitting can easily be extended to include telluric absorption as well \cite[see also][]{2014ASInC..11...53H}. For this we created model spectra for the Earth's atmosphere with varying abundances for molecules that have absorption features in the wavelength range of \muse\ (H$_2$O, O$_2$, and O$_3$) with the FASCODE \citep{2005JQSRT..91..233C,1992JGR....9715761C} based LBLRTM\footnote{\url{http://rtweb.aer.com/lblrtm_frame.html}} (Line-By-Line Radiative Transfer Model), which has been used before for removing telluric lines, among others, by \cite{2010A&A...524A..11S}. LBLRTM always includes the absorption of water in the spectra, so in order to create, for instance, the O$_2$ spectra, we divided by the spectra for water.

A simple polynomial interpolator is used for deriving the absorption spectra used in the fit. Since we handle all the molecules individually, a fifth order polynomial is fitted to each wavelength point for all the spectra for a given molecule. Evaluating this polynomial with the abundance of the corresponding molecule gives us the absorption spectra $T_n$ that are multiplied to get the final telluric spectrum, which finally is convolved with a kernel $C$ in order to allow for a line shift and broadening as with the stellar model spectra.

The model spectrum used for fitting the observation now is just the stellar model multiplied with the telluric model:
\begin{equation}
 G = \sum\limits_{k=1}^{K} \omega_k \left[ B_k \ast M_k \right] \cdot C \ast \prod\limits_{n=1}^{N} T_n.
\end{equation}

By adding these terms, we increase the number of free parameters in the fit by N+2, i.e.\ one for each of the $N$ molecules plus line shift $\sigma_\mathrm{tell}$ and broadening $v_\mathrm{rad,tell}$. While N=3 for \muse, we have shown before in \cite{2014ASInC..11...53H} that this method works well for spectra with a wider wavelength range by using more molecules (CO$_2$, N$_2$O, and CH$_4$). By using a simple polynomial for the interpolation, the increase in computation time as compared to a fit without tellurics is almost negligible.

An example for the Fraunhofer A band found in a typical G subgiant from our sample is shown in Fig.~\ref{figure:tellfit}. As one can see, the residuals are of the order of a few per cent of the flux, which is of the same order as the noise at a S/N of about 100. Marked in the residuals with arrows are the positions of of two \ion{K}{i} absorption lines from the interstellar medium (ISM). A detailed discussion on ISM and diffuse interstellar bands (DIBs) will be presented in another part of this series (Wendt et al., in prep.).

This method yields one telluric spectrum per extracted stellar spectrum, i.e.\ thousands of telluric spectra per pointing that can be averaged to obtain the most likely one. However, tests showed that fitting the extracted stellar spectra with this fixed telluric component does neither improve the results nor reduce the computation time significantly.

\section{Results}
\label{sec:results}
\begin{figure}[t!]
  \includegraphics{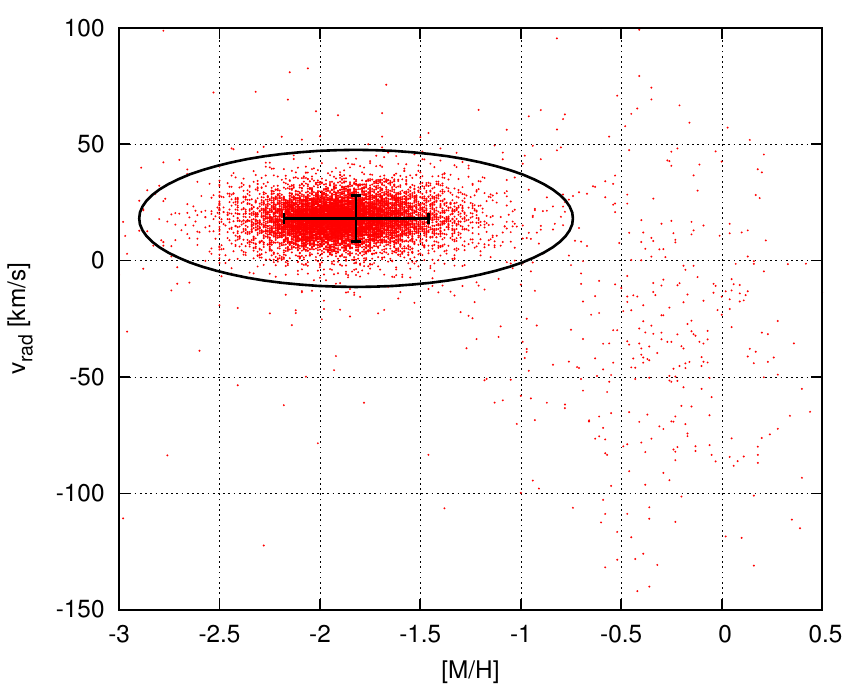}
  \caption{Measured radial velocities \vrad\ over metallicities \mh. A kappa-sigma clipping at 3$\sigma$ (indicated by the overplotted ellipse) on this data is used for determining membership of stars to the cluster.}
  \label{figure:members}
\end{figure}

\subsection{Uncertainties}
For some of the stars in the sample we have more than one spectrum from overlapping pointings or from multiple visits. These independent observations allow us to investigate the quality of the statistical uncertainties returned by the Levenberg-Marquardt optimisation. Having pairs $X_1$ and $X_2$ of results for the same star and their respective uncertainties $\sigma_{X_1}$ and $\sigma_{X_2}$, we can calculate the normalised error $\delta_X$ as
\begin{equation}
 \delta_X = (X_1 - X_2) / \sqrt{\sigma_{X_1}^2 + \sigma_{X_2}^2}.
\end{equation}

Although in general $\delta_X$ is distributed in a Student-t distribution, for small uncertainties on the uncertainties it can be described using a normal distribution, which should have a standard deviation of 1 in the case of accurate uncertainties. Our results yielded 1.50 for \teff\ and 1.24 for \mh, so the formal fitting errors underestimate the real statistical uncertainties by $\sim$50\% and $\sim$24\% respectively. 
\begin{table}
  \caption{Average formal uncertainties from full-spectrum fit for \teff\ and \feh\ in given S/N intervals, scaled by the factor determined by the analysis of the statistical errors. Each bin contains the results of about 2\,100 spectra.}
  \label{table:uncertainties}
  \centering                                      %
  \begin{tabular}{ccc}
    \hline \hline
    S/N interval& $\left< \sigma_{\teff} \right>$ [K]	& $\left< \sigma_{\feh} \right>$ [dex] \\ \hline
    10 -- 14	& 231	& 0.43 \\
    14 -- 20	& 158	& 0.22 \\
    20 -- 28	& 100	& 0.16 \\
    28 -- 41	& 60	& 0.13 \\
    >41 	& 34	& 0.07 \\ \hline
  \end{tabular}
\end{table}

A list of scaled uncertainties from the full-spectrum fit for several S/N intervals is given in Table~\ref{table:uncertainties}. So for a typical spectrum from a \muse\ observation with a S/N of 20, we can expect to derive the effective temperature with an uncertainty of $\sim$100\,K and the metallicity will be accurate to 0.16\,dex. 

\begin{figure*}
  \includegraphics[width=18cm]{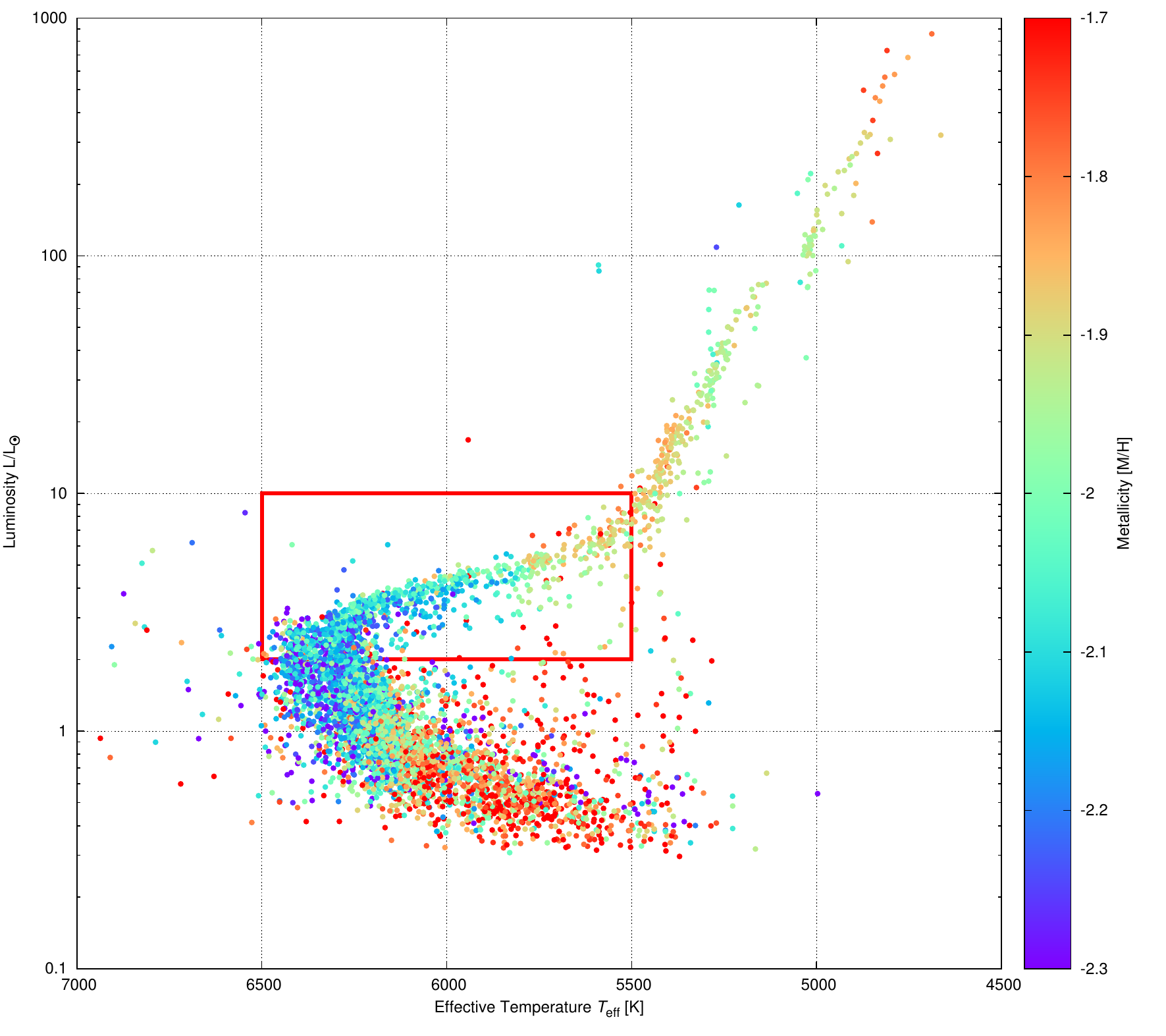}
  \caption{HRD of \object{NGC~6397}, with luminosities plotted as a function of effective temperatures, colour-coded by metallicity. The assumed reddening is E(B-V)=0.18 with a distance modulus of 12.13\,mag. The area on the giant branch marked with a red rectangle will be discussed in Sect.~\ref{sect:metallicities}.}
  \label{figure:hrd}
\end{figure*}
These are of course statistical errors only. We excluded that there are strong systematic errors from our full-spectrum fits caused by the used spectral library by fitting some spectra against \texttt{cflib} 
\citep{2004ApJS..152..251V} and \texttt{ELODIE} \citep{2004astro.ph..9214P}, which both yielded comparable results. As stated before, we also tested our spectral library directly against other libraries, which also did not show any systematics. However, some deviations are expected to be caused by NLTE effects which will be discussed in more detail in Sect.~\ref{sec:nlte}.

\subsection{Radial velocities}
\label{section:vrad}
\begin{figure*}
  \includegraphics[width=18cm]{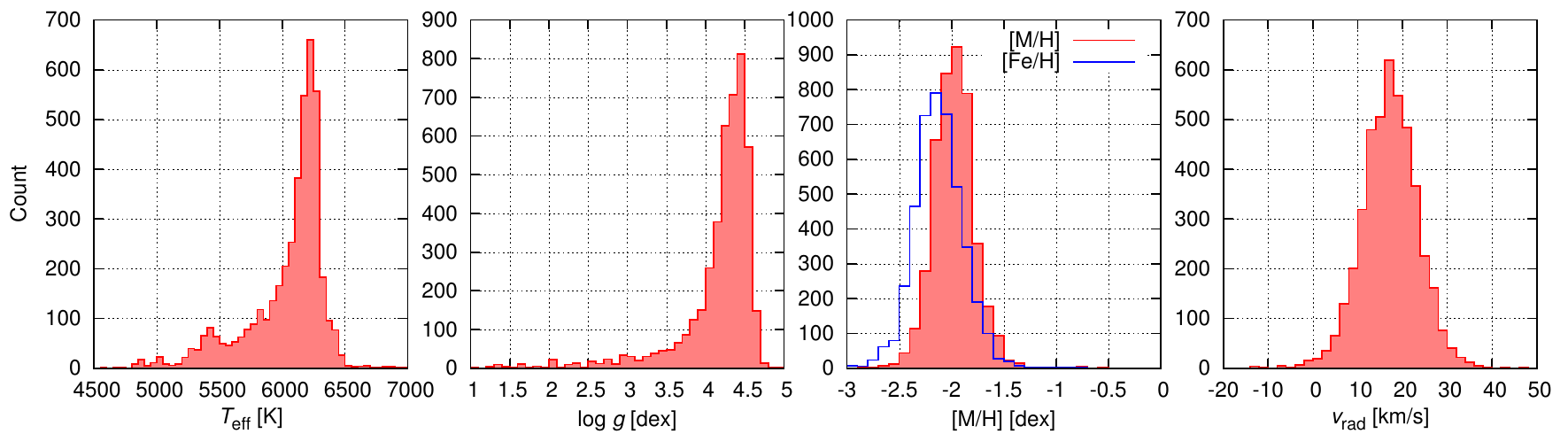}
  \caption{Distributions of the effective temperature \teff, the surface gravity \logg, the metallicity \mh, and the radial velocity \vrad\ as determined in the fits.}
  \label{figure:distributions}
\end{figure*}
All the following results have been obtained for members of the globular cluster only. We defined membership using a simple $3\sigma$ clipping in \vrad-\mh\ space as shown in Fig.~\ref{figure:members}. An implementation of the more elaborate method as described by \cite{2009AJ....137.3109W} is in progress. Since the results described in this publication do not depend critically on determination of membership, the application of that technique will be discussed in \citetalias{2015A&A...subm....K}. With the current method, we determined 7\,142 ($\sim$56\,\%) of the observed stars to be members of \object{NGC~6397}.

The distribution of all spectra with S/N>20 is plotted in the right hand panel of Fig.~\ref{figure:distributions}. We measure a mean radial velocity of \vrad=17.84$\pm$0.07\,km\,s$^{-1}$ ($\sigma$=6.29\,km\,s$^{-1}$), which is a little smaller than the values given in previous studies, but still in good agreement:
\begin{itemize}
 \item $\vrad=18.26 \pm 0.09$\,km\,s$^{-1}$ by \cite{2006A&A...456..517M},
 \item $\vrad=18.1 \pm 0.3$\,km\,s$^{-1}$ by \cite{2008A&A...490..777L},
 \item $\vrad=22.75 \pm 7.1$\,km\,s$^{-1}$ by \cite{2009A&A...505..139C},
 \item $\vrad=20.3 \pm 0.3$\,km\,s$^{-1}$ by \cite{2012ApJ...754...91L}.
\end{itemize}

Another parameter coming from the full-spectrum fit is the radial velocity shift of the telluric spectrum. This can be used for investigating the quality of the velocity results, since it should be close to zero with only a small scatter. In order to achieve this, we had to subtract the barycentric correction, which is done automatically by the pipeline. From our results we get $v_\mathrm{tell}=0.4$\,km\,s$^{-1}$ with a standard deviation of $\sigma_\mathrm{tell}=2.0$\,km\,s$^{-1}$s. Both is consistent with the expected accuracy of the measurements. The little offset seen in the tellurics is most likely caused by a slight asymmetry of the \muse\ LSF that we have not yet modelled correctly.

A more in-depth analysis of the cluster dynamics including a discussion of the corresponding uncertainties will be presented in Paper~II.
\begin{figure}[t!]
  \includegraphics{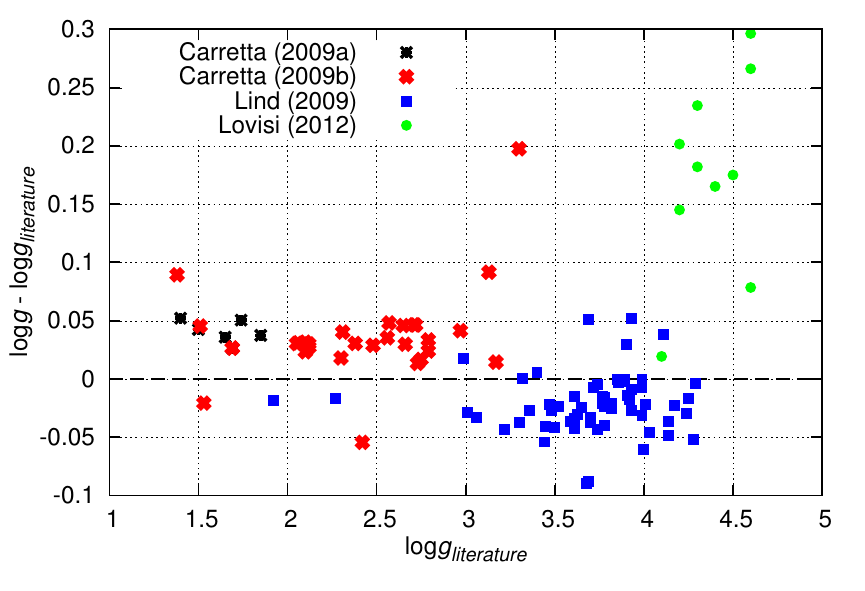}
  \caption{Comparison of our results for \logg\ for single stars that have been analysed before in other studies, all obtained photometrically. The values with the largest deviations by \cite{2012ApJ...754...91L} are mainly those of blue stragglers that we currently cannot fit properly.}
  \label{figure:logg_lit}
\end{figure}

\subsection{Effective temperatures and photometric surface gravities}
We can easily determine absolute V magnitudes from our spectra using a reddening of E(B-V)=0.18 and a distance modulus of 12.13\,mag \cite[see][]{1998AJ....116.2929R}. Applying the polynomial for the bolometric correction from \cite{1996ApJ...469..355F} and the corrections in \cite{2010AJ....140.1158T}, those V magnitudes can be converted into luminosities. Together with the effective temperatures from the full-spectrum fit we can plot an HRD, which is shown in Fig.~\ref{figure:hrd}, colour-coded with the metallicities derived from the fit. Plotted are the results for all member stars with spectra with S/N>20, i.e.\ a total of 3\,870 stars (54\% of the cluster members). To our knowledge, a HRD as detailed as this has never been presented before for a single globular cluster from spectroscopic data. Note that all horizontal branch stars are missing in the HRD, since we currently do not have matching templates for fitting their spectra. 

A comparison of our photometrically obtained surface gravities for single stars with values from the literature is plotted in Fig.~\ref{figure:logg_lit}. Since most of the presented studies -- some of them with a rather large sample -- concentrated on outer fields of the cluster, we could match only a small number of stars to our data. The matching itself has been done by comparing RA/Dec coordinates and magnitudes. 

Although using the FLAMES spectrograph at the VLT, all the discussed studies determined \teff\ and \logg\ photometrically. \cite{2012ApJ...754...91L} used a method similar to the isochrone comparison that we use for our initial guesses, deriving values for \teff\ and \logg\ from theoretical stellar models from the BaSTI database \citep{2006ApJ...642..797P}. In order to be able to use this procedure on blue straggler stars, they compared those to another set of isochrones with different ages. The other three studies \citep{2009A&A...505..139C,2009A&A...505..117C,2009A&A...503..545L} utilised a calibration based on colour indices as described by \cite{1999A&AS..140..261A} that works for red giant stars only.

Since we fix the surface gravities to the values obtained photometrically, we can compare those values directly to the literature (see Fig.~\ref{figure:logg_lit}), which yields an offset of $\delta_{\logg}=0.01$ with a scatter of $0.07$. Ignoring the results from \cite{2012ApJ...754...91L} that mainly consists of blue stragglers -- which we currently do not fit properly --, this number goes down to $\delta_{\logg}=0.00\pm0.04$. 

\begin{figure}[t!]
  \includegraphics{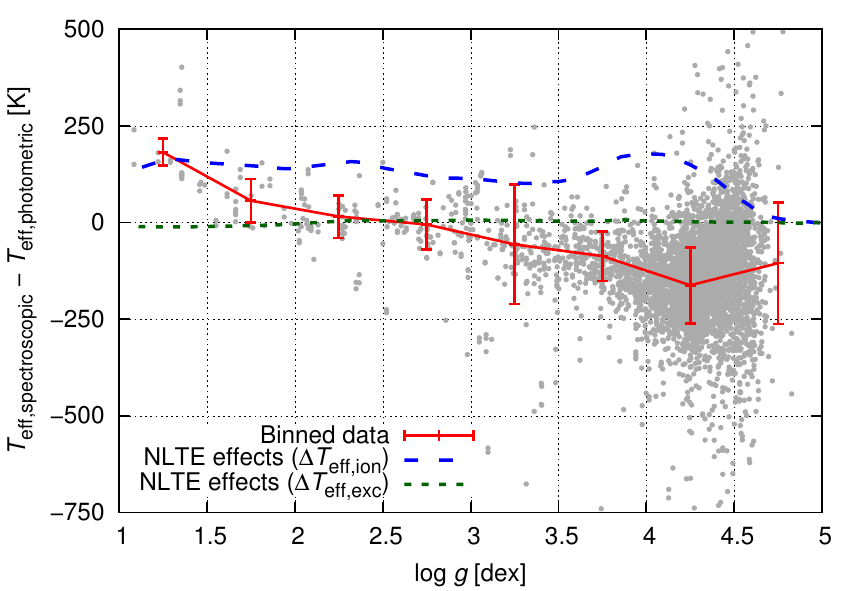}
  \caption{Comparison of our photometric and spectroscopic results for \teff\ as a function of \logg. The red error bars indicate the mean and the standard deviation of the discrepancies in 0.5\,dex wide bins. Overplotted in blue and green are predicted offsets due to neglected NLTE effects as shown in \cite{2012MNRAS.427...50L}. See text for details.}
  \label{figure:teff}
\end{figure}

A comparison of the photometric values for \teff\ from the same publications with our results yields a rather small $\delta_{\teff}=-68.34\pm62.63$\,K. On the other hand it is still unclear whether our photometric or spectroscopic results for \teff, which are compared in Fig.~\ref{figure:teff}, are more accurate. Although the differences are usually small with $\delta T<200$\,K, we see some systematic discrepancies. 

One possible explanation might be NLTE effects that will be discussed in Sect.~\ref{sec:nlte}. However, more likely seems a simple mismatch of the isochrone. Since the comparison with an isochrone was always meant for obtaining initial guesses only, we pick the isochrone manually by eye. Tests showed that the error inflicted on the derived surface gravities is small (of the order of $\Delta\logg\lesssim0.1$\,dex), but that the deviations in colour can be quite large -- especially at the TO and on the RGB -- which directly affects the effective temperature. But since the value for \teff\ is used as initial guess only, this discrepancy is acceptable.

\subsection{Metallicities}
\label{sect:metallicities}
As discussed before, we measure two different kinds of \emph{metallicities}: If \afe\ is fixed to zero in the fit, we denote the metallicity \mh, and \feh\ otherwise. The distribution for both is shown in  Fig.~\ref{figure:distributions}. In the following, we will mainly concentrate on \mh, which we expect to be more accurate.

The mean values were fitted as \mh=-1.962$\pm$0.002 ($\sigma$=0.181) and \feh=-2.120$\pm$0.002 ($\sigma$=0.214). We employed a maximum likelihood technique as described by \cite{1993ASPC...50..357P} for determining the intrinsic dispersion, which is about 0.14\,dex for both quantities. This also increases the mean values slightly by about 0.02\,dex, since usually results for stars with lower metallicities have larger uncertainties.

As mentioned before, we found some discrepancies between the values for \teff\ obtained spectroscopically and from the comparison with an isochrone (see Fig.~\ref{figure:teff}). Since shifting \teff\ also changes \mh, we determined the metallicity for the central field again with both \teff\ and \logg\ fixed to the values derived from the isochrone. As a result we found the metallicity to be shifted slightly, with a maximum offset of $\mh_\textrm{spec}-\mh_\textrm{phot}=-0.07$\,dex at the turnoff, where also the difference in \teff\ is largest.

Most recent studies of \object{NGC~6397} based their metallicity values on equivalent width measurements of \ion{Fe}{i} and \ion{Fe}{ii} lines, which correspond to our \feh\ values. For stars between the TO point and the tip of the RGB, \cite{2009A&A...503..545L} found \feh=-2.10. A similar value of \feh=-2.12$\pm$0.01 was found by \cite{2012ApJ...754...91L} for stars near the TO. All these results are in good agreement with our value for \feh. On the other hand, \cite{2009A&A...505..117C} reported \feh=-1.993$\pm$0.003 for stars along the RGB, which differs significantly from the other values.

\begin{figure}[t!]
  \includegraphics{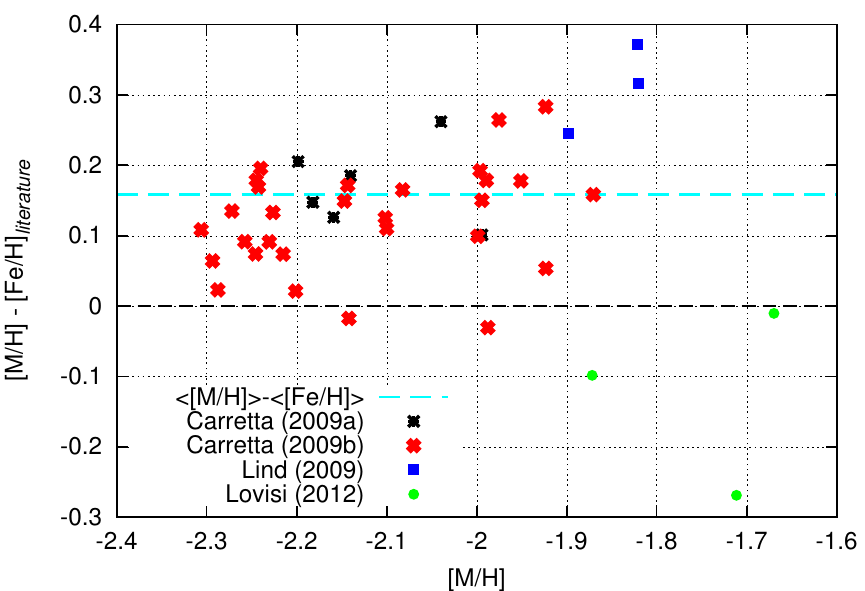}
  \caption{Comparison of results for \mh\ with those from previous high-resolution studies. The dashed cyan line indicates the offset between our mean results for \feh\ and \mh\ of 0.16\,dex.}
  \label{figure:feh_lit}
\end{figure}

A comparison of results for the metallicity of individual stars with values from the literature is plotted in Fig.~\ref{figure:feh_lit}. We present our results for \mh, which show a deviation from the literature values of $\delta=0.12\pm0.10$\,dex. While the offset is in good agreement with the difference between our mean metallicity results, i.e. $\left<\mh\right>-\left<\feh\right>$=0.16\,dex (indicated as a dashed cyan line), the small standard deviation confirms our goal of determining metallicities with an accuracy of 0.1\,dex or better.

\begin{figure}[t!]
  \includegraphics{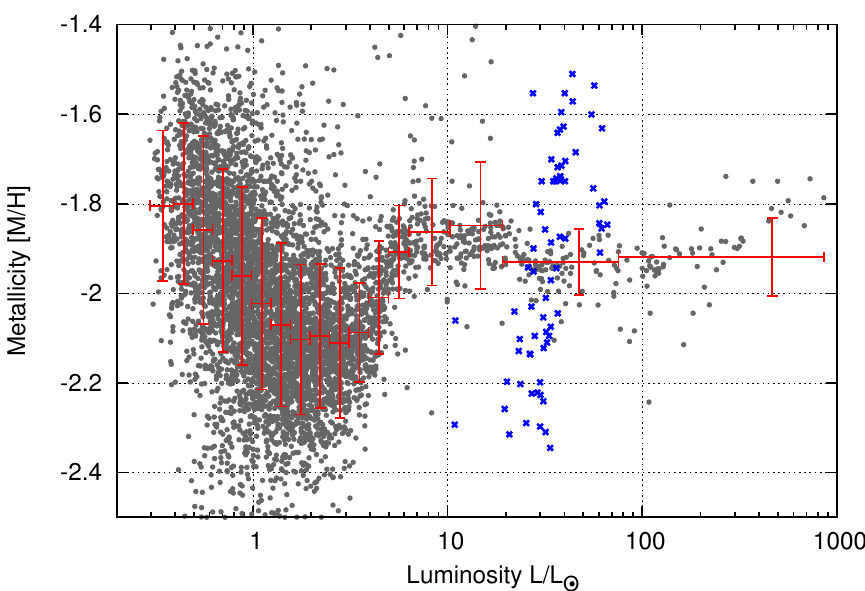}
  \caption{Results for metallicity \mh\ plotted over luminosity $L$. The bins for the error bars contain at least 100 stars each (except for the last one). Marked with blue crosses and excluded from the binning are the HB stars.}
  \label{figure:lum_mh}
\end{figure}

In the HRD (Fig.~\ref{figure:hrd}) we see a variation in metallicity, which gets more pronounced when plotting \mh\ over luminosity (see Fig.~\ref{figure:lum_mh}, for all spectra with S/N>20). Starting on the main sequence with a metallicity of $\mh\approx-1.8$\,dex, this value decreases until it hits a minimum at the turnoff point with $\mh\approx-2.1$\,dex. Afterwards it is increasing again and shows some minor oscillation within the uncertainties around $\mh\approx-1.9$\,dex. Altogether we see the largest variations on the MS and the SGB, while those on the RGB are significantly smaller.

Figure~\ref{figure:teff_feh} concentrates on the increase of metallicity on the RGB after this minimum (around 5\,750\,K). Plotted is the metallicity as a function of effective temperature for all spectra with S/N>50 and \logg<4.4, i.e.\ for all stars at the TO and on the red giant branch (RGB). Again, one can see a significant trend, that also does not disappear with the photometric \teff\ described above.

A similar phenomenon has been reported by \cite{2007ApJ...671..402K} and \cite{2012ApJ...753...48N}, who suggested that extremely metal-poor stars are heavily affected by atomic diffusion. In that scenario, heavy elements in stars on the main sequence sink deeper into the atmosphere and slowly disappear from the spectrum. When reaching the turn-off, convection kicks in and transports the heavy elements back up into the higher atmosphere -- so we should observe an increase of metallicity for stars evolving on the RGB. Model calculations have been done by \cite{2007ApJ...671..402K}, showing the expected trend. For the iron abundance those models (Olivier Richard, priv.\ comm., blue line in Fig.~\ref{figure:teff_feh}) predict a change of $\Delta\feh=0.12\,\mathrm{dex}$ around 5\,750\,K, which is a little shallower than what we see in our data (red line in the same Figure), but still gives a better fit than a straight line.

\cite{2009A&A...501..519B} have also found discrepancies with the abundances for C, Sc, Ti, Cr, Mn and Co being $\sim$0.2\,dex larger in TO stars than in giants. Since they see a physical explanation for the abundance shift in C only, they assume that this effect is not real and caused by shortcomings in their analysis.

Atomic diffusion would also explain the decrease in metallicity along the main sequence towards the turn-off as seen in the HRD (see Figs.~\ref{figure:hrd} and \ref{figure:lum_mh}). Unfortunately, we see a trend that is larger than predicted, so we believe this to be partly caused by the low S/N of the spectra on the main sequence.

When assuming that there is atomic diffusion happening in the atmospheres of the stars, giving a mean metallicity for the cluster is somewhat problematic: the majority of stars in our sample is near the TO, where the change in metallicity caused by diffusion is supposed to be strongest -- this would result in a lower mean value. Therefore we measured the metallicity for stars near the tip of the RGB with \logg<3 and \teff<5\,400\,K, where the change in metallicity due to diffusion is negligible. The mean S/N of the spectra from those stars is $\sim$150 and we get a mean metallicity of $\left< \mh \right>_{RGB\, tip}~=~-1.875 \pm 0.003$\,dex. With the offset discussed before this gives $\left< \feh \right>_{RGB\, tip}~=~-2.04$\,dex, which is still in good agreement with literature values.

\subsection{NLTE effects}
\label{sec:nlte}
We are fully aware that analysing spectra of low-metallicity stars like in our sample with LTE model spectra has some shortcomings. An overview of this topic is given by \cite{2005ARA&A..43..481A} and more recently by \cite{2014arXiv1403.3088B} for the FGK regime.

The possible errors that might occur when running an analysis with LTE model spectra were discussed in \cite{2012MNRAS.427...50L}, based on NLTE models presented in 
\cite{2012MNRAS.427...27B}. The authors determined the offsets in stellar parameters obtained from LTE and NLTE analyses. For \teff\ two different values are given, depending on the type of measurement, i.e.\ derived from either the ionisation balance of high-excitation \ion{Fe}{i} and \ion{Fe}{ii} lines ($\Delta T_\mathrm{eff,ion}$) or from their excitation balance ($\Delta T_\mathrm{eff,exc}$). Under the assumption that our $T_\mathrm{eff, spectroscopic}$ is the \textit{real} effective temperature, we calculated both corrections as shown as green and blue dashed lines in Fig.~\ref{figure:teff}. While $\Delta T_\mathrm{eff,exc}$ can be ignored completely for our purposes ($\lesssim 10$\,K), $\Delta T_\mathrm{eff,ion}$ is almost constant on the RGB and only shows some significant variations at the TO and on the MS. Although there the correction has a similar absolute value as the discrepancy between our photometric and spectroscopic results, it has the opposite sign. So although errors due to the LTE analysis may be present, they do not explain the observed deviations, making an isochrone mismatch (as discussed before) more likely.
\begin{figure}[t!]
  \includegraphics{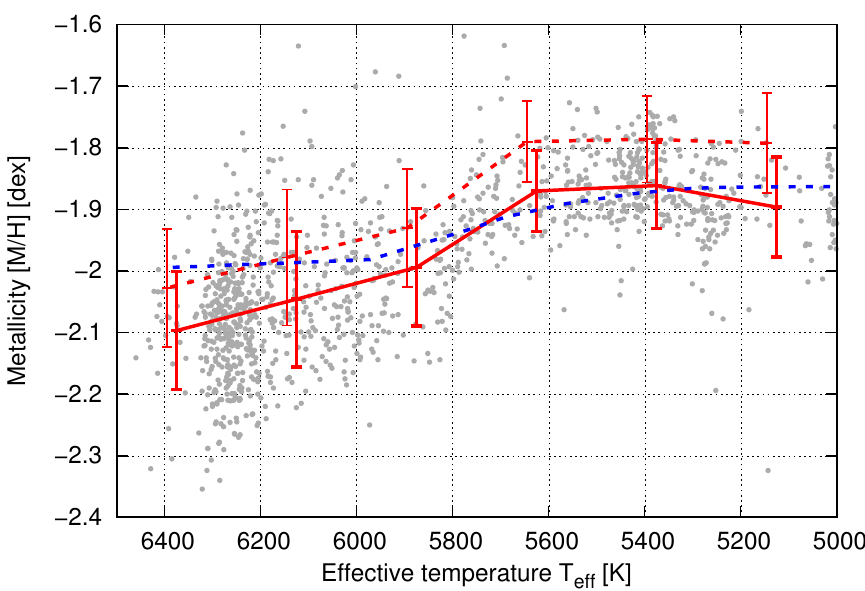}
  \caption{Metallicity trend with effective temperature for all spectra with \logg<4.4 (i.e.\ stars between the TO and the tip of the RGB) and S/N>50. Overplotted with a solid red line are the mean values and standard deviations of the metallicity in 250\,K wide bins. The same is plotted again with a dashed red line including a correction \citep{2012MNRAS.427...50L} for NLTE effects (with the same uncertainties). Overplotted in blue is an isochrone including atomic diffusion (Richard, priv.comm.). See text for details.}
  \label{figure:teff_feh}
\end{figure}

It can be questioned whether these NLTE corrections apply to our results with their full amount. Instead of measuring \ion{Fe}{i} and \ion{Fe}{ii} lines only, we fit a model to the full observed spectrum. A closer look to the NLTE effects caused by other elements can be obtained when using corrections from \cite{2014arXiv1403.3088B}. The most prominent lines in our spectra (besides those from hydrogen) are created by Mg, Ca, and Fe, so Table~\ref{table:nlte_elements} lists the NLTE corrections for these elements for the three stars \object{HD~84937} (TO), \object{HD~140283} (SGB), and \object{HD~122563} (RGB). As one can see, the variation of the corrections between TO and RGB is at most 0.05\,dex, i.e.\ well within our targeted accuracies. Nevertheless, this correction would make the systematic offset between TO and RGB slightly larger.

NLTE effects might also be responsible for the metallicity trend discussed in Sect.~\ref{sect:metallicities}. In Fig.~\ref{figure:teff_feh} we also show the measured metallicities with applied NLTE corrections from \cite{2012MNRAS.427...50L} as dashed red line. As one can see this results in a shift with almost constant offset of $0.06-0.08$\,dex, with a maximum of $\sim0.1$\,dex in the coolest bin. The NLTE corrections for other elements provided by \cite{2014arXiv1403.3088B} also show only small variations between TO and RGB and would just shift the metallicity for all stars by $\sim0.15$\,dex (Mg) and $\sim0.20$\,dex (Ca) respectively.

Altogether the deviations caused by NLTE effects seem to be almost constant (within the uncertainties) for all stars within our sample. Therefore, it is unlikely that NLTE effects alone can explain the metallicity trend that we observe on the giant branch.

\subsection{sdO star ROB 162}
\label{sec:sdo}
The massive multiplexing capabilities of \muse\ offer the possibility of finding peculiar objects. One such object is ROB 162 (\object{GSC~08729-01093}), which can easily be identified in the CMD (see Fig.~\ref{figure:cmd}) as being a little blueward of the horizontal branch. It was identified by \cite{1986A&A...169..244H} as a post-AGB star with 0.5-0.55\,$M_\sun$ and they assume it being the central star of a planetary nebula. However, no nebula was found and we also cannot detect any emission features in the \muse\ data, leaving the explanation that it is already dispersed or too faint. Using NLTE model atmospheres they found \teff=52\,000\,K, \logg=4.5, and a solar helium abundance.

\begin{figure}[t!]
  \includegraphics{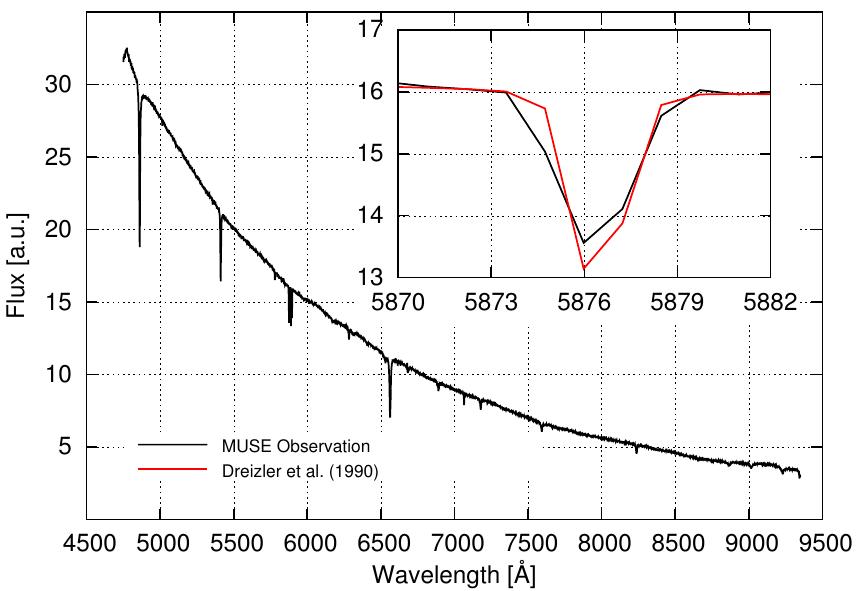}
  \caption{Observed spectrum of sdO star ROB 162, in which the telluric lines have been removed. Overplotted in red in the inset is a synthetic spectra from \cite{1990A&A...235..234D} with \logg=4.5, \teff=52,000\,K, and solar Helium abundances.}
  \label{figure:sdo}
\end{figure}

The observed spectrum of ROB 162 is plotted in Fig.~\ref{figure:sdo}. Overplotted is a TMAP\footnote{\url{http://www.uni-tuebingen.de/de/41621}} spectrum with \teff=52,000\,K, \logg=4.5  and solar Helium abundances by \cite{1990A&A...235..234D} with updates described in \cite{1999JCoAM.109...65W}.
\begin{table}
  \caption{NLTE corrections for single elements as taken from \cite{2014arXiv1403.3088B}.}
  \label{table:nlte_elements}
  \centering                                      %
  \begin{tabular}{lcccc}
    \hline \hline
    Star      & Type & Mg [dex] & Ca [dex] & Fe [dex] \\ \hline
    HD 84937  & TO   & 0.12     & 0.18     & 0.06 \\
    HD 140283 & SGB  & 0.15     & 0.22     & 0.06 \\
    HD 122563 & RGB  & 0.17     & 0.20     & 0.09 \\ \hline
  \end{tabular}
\end{table}

Since there are no spectra of post-AGB stars in our \phx\ library, the general pipeline fit determined ROB 162 to be a very metal poor (\feh=$-3.4$) F type dwarf, which matches the hydrogen lines surprisingly well -- of course without showing any He lines. The radial velocity resulting from this is \vrad=21$\pm$2\,km\,s$^{-1}$. When fitting with the TMAP spectra we obtain 29$\pm$1\,km\,s$^{-1}$s. This difference is caused by the \ion{He}{ii} absorption lines close to $H\alpha$ and $H\beta$, which make their profiles slightly asymmetric. Therefore, the TMAP result should be the more trustworthy value, which can be compared to the radial velocity \vrad=28$\pm$5\,km\,s$^{-1}$ found by \cite{1986A&A...169..244H}.

This result is only $2\sigma$ off the mean velocity of the cluster, so in combination with its position in the CMD near the HB and its proximity to the cluster centre ($\sim$2\,arcmin, well within the 2.90\,arcmin half-light radius), we conclude that ROB 162 really is a member of \object{NGC~6397}.

\subsection{Possible pulsar companion}
\label{sec:pulsar}
Another interesting object is \object{Cl*~NGC~6397~SAW~V31}, for which we have two observations 22 hours apart. Our analysis identifies it as K sub-giant with \logg$\approx$3.6 and \teff$\approx$4\,200\,K. It was fitted with a very low metallicity with \mh$\approx-3$.

\begin{figure}[t!]
  \includegraphics{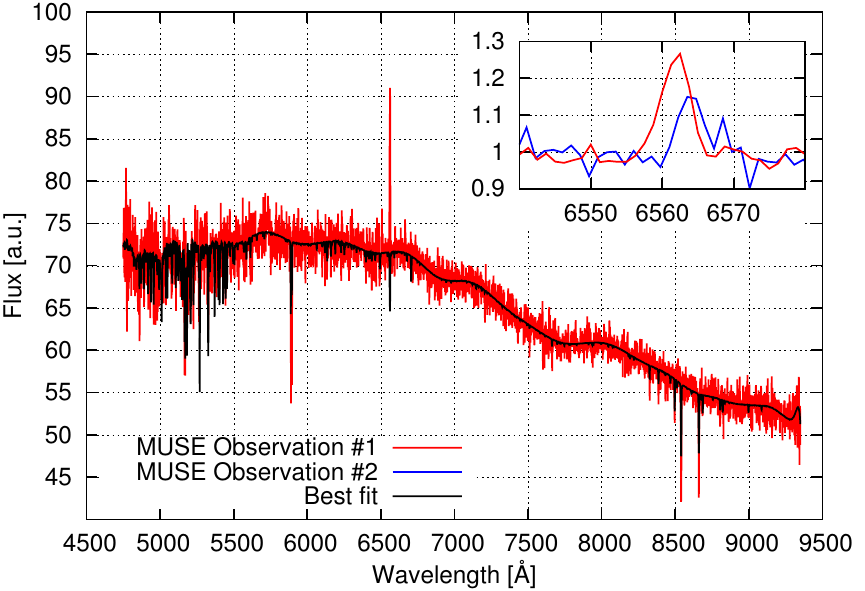}
  \caption{One observation of the possible pulsar companion V31 as observed with \muse. Overplotted is the best fit, identifying it as a K sub-giant, with the residuals shown below. The inset zooms into the region around $H\alpha$, showing the line in emission in both observed epochs and significantly shifted to each other.}
  \label{figure:pulsar}
\end{figure}

Figure~\ref{figure:pulsar} shows the full-spectrum of one of the observations, where one can clearly identify the $H\alpha$ emission line, which is also shown in the inset for both obtained spectra. In there one can see a large radial velocity shift of $\Delta \vrad=145.9$\,km\,s$^{-1}$.

For this object \cite{2006MNRAS.365..548K} observed a periodicity of $\sim$1.3 days, possibly due to ellipsoidal variations. They also connected it to the hard X-ray source U18, which was discovered by \cite{2001ApJ...563L..53G} and identified either as a BY Dra type variable or a millisecond pulsar. This connection, however, was challenged by \cite{2010ApJ...709..241B}, who claim the two objects to be too far apart.

A radial velocity shift as large as the one we have discovered can easily be caused by the motion of V31 within a binary system. If the companion is indeed a pulsar, we can furthermore speculate that our K sub-giant may have a hot day and a cooler night side which could explain both the reported variability and the strong variation in the strength of the $H\alpha$ line in our spectra. With the given period and velocity this yields a lower limit for the total mass of $\sim$0.4$M_\sun$ in the system, consistent with our suggestion.

\section{Conclusions \& Outlook}
\label{sec:conclusions}
We demonstrated how the use of an IFS like \muse\ dramatically increases the number of globular cluster stars available for spectroscopic analyses. Instead of a pre-selected list of just a few hundred spectra per observation campaign, we can obtain thousands of medium-resolution spectra of individual stars per exposure.

Although obtained with a spectral resolution lower than in most other studies, we showed that our spectra are still sufficient for investigating both the atmospheric parameters and the kinematics of the stars. While the results for a single spectrum might be less accurate, the huge amount of data allows for new statistical approaches, resulting in uncertainties not larger than in high-resolution studies. On the other hand the broad wavelength range of \muse\ allows us to determine atmospheric parameters like \teff\ and \mh\ spectroscopically. In principle this should also be possible for \logg, but there we still suffer from our incomplete knowledge of the LSF that needs to be further investigated, resulting in the surface gravity currently being determined photometrically. Overall, we can reproduce results from high-resolution spectroscopic studies very well.

In this paper we presented spectroscopic observations of 12\,307 stars of which we determined 7\,142 to be members of the globular cluster \object{NGC~6397}, most of them (87\%) below the TO. For 3\,870 of those stars we obtained spectra with S/N>20, from which we derived atmospheric parameters and constructed a HRD. To our knowledge, this is the most comprehensive HRD of a globular cluster obtained from spectroscopy so far.

The large number of stars in our sample allows us to investigate problems that before have only been explored using a few stars, namely, for instance, the variation of abundances with effective temperature in the atmospheres of RGB stars. There are indications that the trend that is visible in our data -- a sample of almost 4\,000 spectra of more than 2\,000 stars at the TO and on the RGB -- is real. Extending our methods for measuring the abundances of single elements (like Mg and Ca) and including NLTE effects will allow us to investigate this in more detail in many clusters.

We are aware of some shortcomings in our analysis and discussed the problems of using LTE model spectra and thus ignoring NLTE effects in detail. Fixing the surface gravity to a value obtained photometrically -- which, however, is an approach commonly found in the literature -- is an inconsistency in our analysis, but the effect should be small. Another potential issue are the model spectra used for the analysis: each synthetic spectral library uses other parametrisations for, e.g., the mixing length and macro- and microturbulence, uses other linedata and opacities, and so on. Therefore, this is not a problem affecting our analyses alone. We are confident that our model spectra are suitable for the task at hand.

As for many other globular clusters, there have also been reports of multiple main-sequences found in \object{NGC~6397} \citep{2012ApJ...745...27M,2015A&A...573A..70N}. In this case, however, the effect is so small that we do not expect it to be measurable from \muse\ spectra -- especially with the rather low S/N values we obtained for spectra of main-sequence stars.

Since we are not observing only hand-picked targets as with classical spectroscopic techniques like longslit and MOS, we expect to get a lot of spectra from a variety of peculiar objects, like the sdO and the possible pulsar companion described in this publication. While independent science cases will benefit from these findings, it will also help us to better understand the evolution of globular clusters.

The aspect of kinematics has not been addressed in this publication, although it is a topic of great interest with regard to claims of black holes in the centres of globular clusters over the last decade. A comprehensive discussion in the case of \object{NGC~6397} will be presented in Paper~II.

Altogether we have to emphasise that the observations presented in this publication were taken during the commissioning of \muse, and we were constrained by the overall commissioning goals. However, with the start of regular \muse\ observations, we began using GTO time for a large survey, targeting about 25 Galactic globular clusters over several epochs. Co-adding spectra from multiple visits will allow us to increase the S/N significantly, resulting in better results for single stars, making it, for instance, more feasible to contribute to the research on multiple populations in globular clusters.

\section*{Acknowledgements}
SK and PMW received funding through BMBF Verbundforschung (project MUSE-AO, grant 05A14BAC and 05A14MGA). Based on observations made with the NASA/ESA Hubble Space Telescope, and obtained from the Hubble Legacy Archive, which is a collaboration between the Space Telescope Science Institute (STScI/NASA), the Space Telescope European Coordinating Facility (ST-ECF/ESA) and the Canadian Astronomy Data Centre (CADC/NRC/CSA).

\bibliographystyle{aa}
\bibliography{ngc6397_methods}

\end{document}